\documentstyle[twoside,fleqn,espcrcs,epsfig]{article}

\newcommand\as{\alpha_{s}}
\newcommand\yc{y_{\rm c}}

\def\beq{\begin{equation}}
\def\eeq{\end{equation}}
\def\bea{\begin{eqnarray}}
\def\eea{\end{eqnarray}}

\newcommand{\permil}{\raisebox{0.5ex}{\tiny $0$}$\!/$\raisebox{-0.3ex}{\tiny $\! 00$}{\normalsize \hspace{1ex}}}

\def\cpc#1#2#3{Computer Phys.\ Comm.\ #1 (19#3) #2}

\def\np#1#2#3{Nucl.\ Phys.\ B#1 (19#3) #2}
\def\pl#1#2#3{Phys.\ Lett.\ B#1 (19#3) #2}

\def\prl#1#2#3{Phys.\ Rev.\ Lett.\ #1 (19#3) #2}

\def\zp#1#2#3{Zeit.\ Phys.\ C#1 (19#3) #2}

\begin{document}

\hyphenation{author another created financial paper re-commend-ed}



\pagestyle{empty}

\twocolumn[
{\Large Experimental determination of $m_b(M_Z)$ in DELPHI} 
\vspace*{0.5cm} \\
S. Mart\'{\i} i Garc\'{\i}a$^a$, J. Fuster$^b$ and S. Cabrera$^b$
\vspace*{0.5cm} \\
$^a$~University  of Liverpool, Oxford  St.,  Liverpool L69 7ZE, United
Kingdom, e-mail: s.marti@cern.ch
\vspace*{0.3cm} \\
$^b$~IFIC, centre mixte   CSIC--Universitat de Val\`encia, Avda.  Dr.  
Moliner  50,\\   E--46100   Burjassot,   Val\`encia,   Spain,  e-mail:
fuster@evalo1.ific.uv.es, cabrera@evalo1.ific.uv.es
\vspace*{0.3cm} \\
{\small  The  running  mass   of the  $b$  quark   as  defined in  the
  $\overline{MS}$ renormalization scheme,  $m_b$, was measured at  the
  $M_Z$ scale using 2.8 million hadronic $Z^0$ decays collected by the
  DELPHI experiment at LEP. The result is
\[
m_b(M_Z) = 2.67 \pm 0.25 ({\rm stat.}) \pm 0.34 ({\rm frag.}) \pm 0.27
({\rm theo.)} {\rm GeV}/c^2
\]
which  differs from   that   obtained at   the $\Upsilon$   scale,  by
$m_b(M_\Upsilon/2)-m_b(M_Z) = 1.49\pm  0.52$   GeV/$c^2$. This
measurement, performed  far from the $b\bar{b}$  production threshold,
provides  the first experimental   observation of  the running of  the
quark masses.}  \vspace*{0.4cm}]

\section{Introduction}

The quark masses are   fundamental parameters  of the  QCD  lagrangian which 
are not predicted by the theory.  Their definition is not unique since  quarks
are not observed  as free particles, but confined inside hadrons.  Therefore, 
the use of dynamical  expressions is mandatory in order  to determine the 
masses of  the quarks.   The perturbative pole mass,  $M_q$, and the running  
mass  in  the  $\overline{MS}$ scheme, $m_q$, are two definitions among the 
most frequently used.

In the renormalization procedure of QCD, once the divergent part is 
subtracted, we  deal with  finite quantities for the strong coupling constant,
$\as$,  and  the quark   masses,  $m_q$.   However,  these quantities are not
fixed parameters, but their actual value depends on the energy scale of the 
process being   considered.  In other words, $\as$ and $m_b$ exhibit a 
running property.

The  running of $\as$  has been verified  many times by different experiments 
(for  a review see  \cite{bethke,salva_dis97}).   Although the running of the
quark masses is  as basic to  QCD as that of $\as$, it has never been 
experimentally  tested before.  The empirical determination of the quark 
masses at different scales must be regarded as a fundamental test of QCD.

Nowadays, the  LEP I data allow  measuring  the $b$ quark  mass at the $M_Z$ 
scale (far away from the  $b\bar{b}$ production threshold), thus testing its 
running.  According  to the  $m_b$ evolution, dictated  by the  
Renormalization Group Equation (RGE), a  change on $m_b$ of $\sim 1.3$ GeV 
is expected when varying the scale  from $M_\Upsilon/2$ up to $M_Z$,  as  
$m_b(M_\Upsilon/2)  \approx   4.16 \pm    0.18$  GeV/$c^2$ \cite{german_qcd97}.

In  $e^+e^-\rightarrow$hadrons, the three-jet rate  is an observable sensitive
to the $b$ mass. When testing  the  universality of $\alpha_s$ by using the 
three-jet rate, it  was observed that due to the mass effects, $b$ quarks 
radiate a 3 to 5\% less gluons than light quarks do \cite{delphi_ab}. So, 
reversing the argument: assuming the $\as$ universality allows measuring 
the $b$ quark mass by studying the three-jet rate.

The former Leading  Order (LO)  calculations  of the three-jet cross section 
in $e^+e^-$   including  mass  terms \cite{tor,val}  are   not appropriate in
order  to evaluate   the $b$  quark  mass, as    these calculations can not 
discern  between $M_b$ and $m_b$.  Recently, Next to Leading Order (NLO) 
expressions for the  three-jet rate in $e^+e^-$ are   available 
\cite{german,aachen,nason}, thus, enabling the measurement of $m_b$.

The observable used  in this  work is:
\[
\begin{array}{rl}
R_3^{b{\ell}}(y_{\rm c})& = \frac{\Gamma_{3j}^{Z^0\rightarrow b\bar{b}g}(y_c)/
                      \Gamma_{tot}^{Z^0\rightarrow b\bar{b}}}
                    {\Gamma_{3j}^{Z^0\rightarrow {\ell}\bar{{\ell}}g}(y_c)/
                      \Gamma_{tot}^{Z^0\rightarrow {\ell}\bar{{\ell}}}} = \\ 
                     & 1 + r_b(\mu) \left( b_0(y_{\rm c},r_b) +
                      \frac{\alpha_s(\mu)}{\pi}b_1(y_{\rm c},r_b) \right)
\end{array}
\]
where       $\Gamma_{3j}^{Z^0\rightarrow        q\bar{q}g}$        and
$\Gamma_{tot}^{Z^0\rightarrow  q\bar{q}}$ denote    the   differential
three-jet and the total cross sections for $b$ and light quarks ($\ell
= u,d,s$), $r_b(\mu)  = (m_b(\mu)/M_Z)^2$.  The coefficients $b_0$ and
$b_1$ are calculated at NLO in reference \cite{german}. $y_{\rm c}$ is
the   jet resolution  parameter and  $\mu$   the characteristic energy
scale.

\section{Data analysis}
The  analysis reported was performed over   a sample consisting of 2.8
million Z$^0$ hadronic decays recorded by the DELPHI detector at LEP
through the years 1992 to 1994.

\subsection{Event  selection}
Those   hadronic events  that satisfied the   flavour tagging (section
\ref{sec:proba}) and the jet reconstruction criteria were retained for
the analysis.
 
The jets of the event  were reconstructed by means  of the Durham  jet
finding algorithm  \cite{durham}.  By choosing the appropriate $y_{\rm
  c}$, it was possible to force  the reconstruction of just three jets
in every  event.  Then, a set of  quality cuts were  applied over each
jet  (minimum  charged  multiplicity,    enough  visible energy,   jet
direction  in  the    barrel   part of  DELPHI    and   overall planar
configuration).  These conditions  had to  be simultaneously fulfilled
by all three jets.

A total of 1,074,860 and 294,509 events  entered in the $\ell$ and $b$
sample respectively.

\subsection{Flavour definition}
The true flavour of the event was defined as that of the quark coupled
to the $Z$ in the  $Z\rightarrow q\bar{q}$ vertex, without prejudice to
the new flavour production that  possibly occurred in the splitting of
a bremsstrahlung gluon (e.g.   $g\rightarrow b\bar{b}$).  This is  the
same  criterion adopted  in \cite{german}, thus   it enables  a direct
comparison of our result with that NLO calculation.

\subsection{Flavour tagging technique \label{sec:proba}}
The signed impact parameter of all charged  particles in the event was
used   to set   up an   algorithm  which enables  the  flavour tagging
\cite{borisov}.   The significance   of each  track   was obtained  by
weighing  its  impact parameter with   its  associated error.  Thus, a
function,    $\cal   P$, was constructed    in order   to estimate the
probability of having all particles compatible with being generated in
the events'  Interaction   Point (IP).  By construction,  light  quark
events had  an uniform distribution  of  $\cal P$.   In the $b\bar{b}$
events, the decay of  long lived B hadrons  led to particles generated
in secondary vertices far away from the  IP, besides, biasing $\cal P$
towards low  values.  Accordingly, $b\bar{b}$  events were selected by
requiring  ${\cal  P}<5\cdot10^{-3}$ and $\ell\bar{\ell}$  with ${\cal
  P}>0.2$.  The purity attained for the $b$ sample is approximately an
85\% and its efficiency a 55\%, while for $\ell$ quarks both are about
the $80\%$.  The  contents on $c\bar{c}$ events  in each tagged sample
($b$ and $\ell$) were estimated to be a 10\% and a 15\% respectively.

On the  contrary to the  $\ell$ flavour,  the $b$  tagging  efficiency
depends on the  number of jets  in the event.   The mean energy of the
$B$ hadrons in a  three-jet event is  smaller than  that in a  two-jet
event. The  same  applies to their   products and due to  the multiple
scattering,  the  error  of  their   impact   parameter  is  larger.   
Consequently, the probability of having  all particles compatible with
being produced in the IP is also larger.

The background of the $b$ sample comes mainly from those $\ell$ events
having either  not identified $V^0$  decays or $\gamma$ conversions in
the  innermost layers of the  detector.  The large impact parameter of
their products pulls $\cal P$ towards low values.

\subsection{Jet rates}
The observed three-jet cross section  of each tagged sample, $q=b$ and
$\ell$, was computed as:
\[
R^{\rm obs}_{3q}  (y_{\rm  c})=  \Gamma^{Z\rightarrow
  q\bar{q}g}_{3j{\rm     -obs}}(y_{\rm    c})/   \Gamma^{Z\rightarrow
  q\bar{q}}_{{\rm tot}{\rm -obs}}
\]

The measured value of our  observable can be  computed simply with the
quotient of the reconstructed three-jet rates:
\[R^{b\ell\rm  -obs}_3 (y_{\rm c}) =
R^{\rm  obs}_{3b}(\yc)/R^{\rm obs}_{3\ell}(\yc)\]
 
This raw value of the observable must be converted into a parton level
one    which     may   be compared      with    the  NLO  calculations
\cite{german,aachen}.  The correction method accounts for the detector
effects,  biases introduced    in   the  flavour tagging   plus    the
hadronization effects.

The contribution   of    each flavour, $R^i_{3q}$,    to  the observed
three-jet cross section is given by:
\[ 
R^{\rm obs}_{3q} = c^b_q \cdot R^b_{3q} + c^c_q \cdot R^c_{3q} + 
c^{\ell}_q \cdot R^\ell_{3q}
\] 
where   $c^i_q$   corresponded  with    the   flavour   contents   for
$i=b,~c,~\ell$ of both  tagged samples.  These factors were  extracted
from the simulation.

The reconstruction and parton   level three-jet rates of  each flavour
and  sample ($R^i_{3q}$  and   $R^{\rm par}_{3q}$   respectively) are
related by:
\[
R^i_{3q}(y_{\rm c}) = f^i_{3q}(\yc)\cdot g_{3i}(\yc)\cdot 
R^{\rm par}_{3i}(\yc) 
\]
where $f^i_{3q}$ corresponds with the factor correcting both: detector
acceptance and tagging effects.  These factors can be deduced from the
modelling of the DELPHI  response to the  hadronic events.  Of course,
data sets corresponding to different years of the detector's operation
must be corrected independently.  The hadronization factors, $g_{3i}$,
can be estimated by comparing the hadron and parton level distribution
of the three-jet rate.

The   impact of the  $c$  quark mass in  our  observable can be safely
neglected.  As mass effects are proportional to $m_q^2$, the effect of
$c$ quarks compared to  $b$'s is $(m_c/m_b)^2  \sim 1/10$ times a 10\%
factor,  due to  the contents on  $c$  flavour of the  tagged samples. 
Hence, the influence of the $c$ mass would be roughly an $1\%$ that of
the $b$, i.e.  a  repercussion below the  0.5\permil in both, numerator
and denominator of the  observable. Furthermore, the ratio reduces the
net effect to negligible levels.  According to  this, the parton level
jet rates  of the $c$  quarks are taken equal  to  those of  the light
quarks: $R^{\rm   par}_{3c}  \equiv  R^{\rm par}_{3\ell}$.    Now, the
measured jet rates can be expressed as:
\[ 
R^{\rm obs}_{3b}(\yc) = A_b(\yc) \cdot R_{3b}^{\rm par}(\yc) + 
B_b(\yc) \cdot R_{3\ell}^{\rm par}(\yc)  
\] 
\[ 
R^{\rm obs}_{3\ell}(\yc) = A_\ell(\yc) \cdot R_{3b}^{\rm par}(\yc) + 
B_\ell(\yc) \cdot R_{3\ell}^{\rm par}(\yc)  
\] 
where $A_q$ and   $B_q$   are a redefinition  of   the
original set  of  parameters:  $c^i_q,~f^i_{3q}$ and  $g_{3i}$.  This
parametrization allows expressing the corrected observable as:
\[
R_3^{b\ell}(\yc) = \frac{R_{3b}^{\rm par}}{R_{3\ell}^{\rm par}} =
\frac{B_b - B_{\ell}\cdot R_3^{b\ell{\rm -obs}}}
{A_{\ell}\cdot R_3^{b\ell{\rm -obs}} - A_b}
\]

\begin{figure}[tb]
\vspace*{-6mm}
\epsfig{file=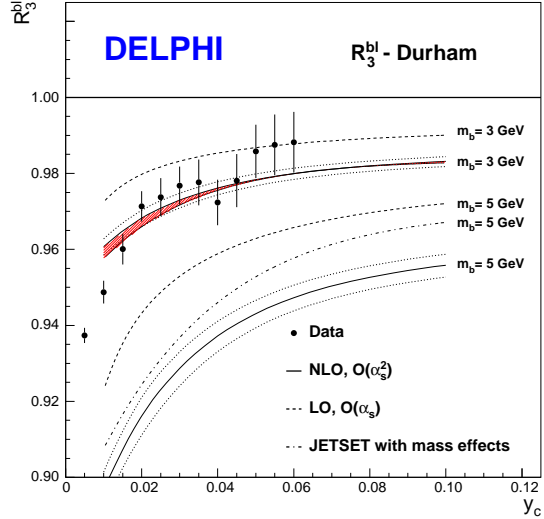,height=79mm}
\caption{\small Experimental values of $R_3^{b\ell}(y_{\rm c})$ compared with
  different QCD predictions at LO and NLO for two different mass hypothesis
  $m_b = 3, 5$ GeV/$c^2$.  The experimental  points are displayed only
  with their statistical error.}
\label{fig:r3bl}
\end{figure}

The total correction applied to  the  raw $R_3^{b\ell}$ was about  the
10\%,  the bulk of   which corresponded to   the detector plus tagging
effects  and   an $\sim 1\%$  to   the  hadronization.  The  corrected
$R_3^{b\ell}$ distribution is   shown in  figure \ref{fig:r3bl}.   The
experimental points are  seen to lay  well below  1.  So, effectively,
$b$ quarks  radiate  less gluons  than light  quarks do.  In  order to
quantify this statement,   the measurement must be  based  on a single
point, since  all data points  are highly correlated.  Large values of
$\yc$  must be  avoided in order  to keep  low  the statistical error. 
Also low values  of $\yc$ should be eluded,  thus limiting the effects
of the higher order terms.  Then,  $\yc=0.02$ is chosen and the result
is:
\[
R_3^{b\ell}(0.02) = 0.971 \pm 0.005 ({\rm stat.}) \pm 0.007 ({\rm frag.})
\]

\subsection{Error estimation \label{sec:syst}}
The correction  of the  measured  $R_3^{bl}$ induced  some  systematic
uncertainties.     The  source of    systematics  considered   are the
following: fragmentation and simulation.

The fragmentation   model  uncertainty has   been   evaluated  by  the
comparison of two models:  string fragmentation (JETSET \cite{jetset})
and  cluster decay  (HERWIG \cite{herwig}).   Huge  samples were  used
(more than  $10^7$ simulated events per  generator  and flavour).  The
adopted value of the $R^{b\ell}_3$  was the average of those extracted
correcting   with both models,   which  differed by   about  1\%.  The
fragmentation  model uncertainty was  taken to  be half of  difference
between them.

The     effect     of      the    main     fragmentation    parameters
($Q_0,~\sigma_q,~\epsilon_b,~a$ and   $b$ in JETSET)  was studied also
with massive  statistics ($10^7$ events   per flavour and  parameter). 
Each  of the above   parameters took the  optimum value  found in  the
DELPHI tuning \cite{delphi_jetset} and  varied by $\pm 2\sigma$.   For
$\yc   >0.01$,   none of  the  uncertainties    due to  the individual
parameters exceeded  the $2$\permil.  The total  error was obtained by
adding     quadratically the  individual   ones,   and  neglecting the
correlation.  It represented roughly a 3\permil.

\begin{figure}[tb]
\epsfig{file=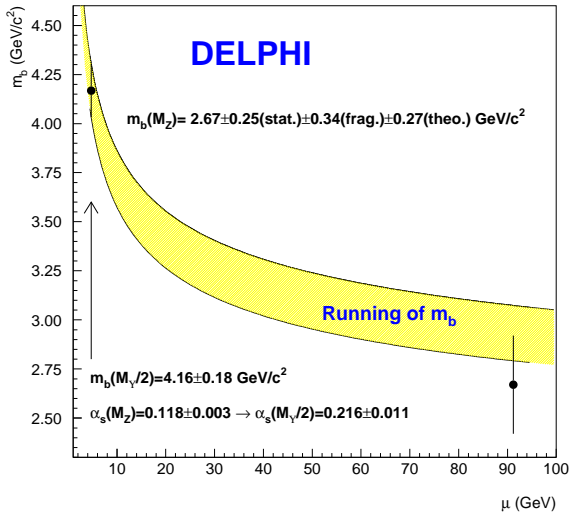,height=68mm}
\caption{\small Experimental result of $m_b(M_Z)$. The DELPHI point at 
  $M_Z$ is displayed only  with its statistical  error.  It is seen to
  be compatible with the running of $m_b$ as predicted by QCD.}
\label{fig:mbmz}
\vspace*{-4mm}
\end{figure}

The defects of the DELPHI detector  modelling when trying to reproduce
the real data were also considered in  the analysis and regarded as an
additional source of error.  The main contribution to this uncertainty
came from the limited statistics   of  the simulated events   ($4\cdot
10^6$) with full DELPHI  simulation.   Thus, the limited knowledge  on
the purity/background factors, $c^i_q$,  of the tagged samples entered
as the simulation error.  This error  has a strong dependence on $\yc$
as it is linked  to the statistics of the  three-jet simulated sample. 
But, for $\yc=0.02$, it represents  a mere 3\permil (already  included
in the statistical error).

\section{The running mass of the $b$ quark}
Using   the DELPHI  result of $R_3^{b\ell}(\yc)$   and  the recent NLO
calculations for exactly the same  observable \cite{german}, the value
of the running mass of the $b$ quark at the $M_Z$ scale is found to be:
\[
m_b(M_Z) = 2.67 \pm 0.50~{\rm GeV}/c^2
\]
where the breakdown of the  error is listed  in figure \ref{fig:mbmz}. 
The theoretical  uncertainty has  two  sources.  The first one  is the
error due to the   QCD scale.  It  has been  estimated by varying  the
scale,  $\mu$, in the range: $0.5\leq  \mu/M_Z \leq 2$, and represents
0.10  GeV$/c^2$.  The second  one is due  to the {\sl mass ambiguity}. 
It accounts for the  uncertainty of expressing  the mass dependence of
$R^{b\ell}_3$  by either using  directly  the running mass, $m_b$,  or
using the  pole mass, $M_b$, and then  run up to $M_Z$  using the RGE. 
Both practices are  licit, but differ on the  treatment of  the higher
order    terms (with   truncated   or    resummed expressions).    The
contribution to the error is 0.25 GeV$/c^2$.

The running of $m_b$ is verified when comparing its value at the $M_Z$
with that at the  $\Upsilon$ scale. The net change  in the $b$ running
mass between both scales is (errors added up quadratically):
\[
m_b(M_\Upsilon/2) - m_b(M_Z) = 1.49 \pm 0.52~{\rm GeV}/c^2
\]

This result represents the  first experimental evidence of the running
property of the $b$ quark mass, in  particular, and of the any fermion
mass, in general.


\end{document}